\documentclass[aps,pre,preprint]{revtex4-1}

\usepackage{epsf}
\usepackage{latexsym}
\usepackage{epsfig}
\usepackage{amsmath}
\usepackage{amsfonts}
\usepackage{amssymb}
\usepackage{graphicx}
\usepackage{color}
\usepackage[usenames,dvipsnames]{xcolor}

\newcommand{\prs}[1]{{\left(#1\right)}}
\newcommand{\col}[1]{{\left[#1\right]}}
\newcommand{\chs}[1]{{\left\{#1\right\}}}

\newcommand{\vc}[1]{{\boldsymbol #1}}

\newcommand{\sC}{{\mathcal{A}}}

\begin{document}

% ================
% Title and Author
% ================
\title{Measuring Complexity through Average Symmetry}
\author{Roberto C. Alamino}
\affiliation{Non-linearity and Complexity Research Group, Aston University, Birmingham B4 7ET, UK}

% ========
% Abstract
% ========
\begin{abstract}
This work introduces a complexity measure which addresses some conflicting issues between existing ones by using a new principle - measuring the \emph{average} amount of symmetry broken by an object. It attributes low (although different) complexity to either deterministic or random homogeneous densities and higher complexity to the intermediate cases. This new measure is easily computable, breaks the coarse graining paradigm and can be straightforwardly generalised, including to continuous cases and general networks. By applying this measure to a series of objects, it is shown that it can be consistently used for both small scale structures with exact symmetry breaking and large scale patterns, for which, differently from similar measures, it consistently discriminates between repetitive patterns, random configurations and self-similar structures.
\end{abstract}

\maketitle

% ============
% Introduction
% ============
\section{Introduction}

Complexity is a cross-disciplinary concept largely used in a number of fields ranging from physics~\cite{Crutchfield12} to social sciences~\cite{Derex13}. Measures of complexity in different systems have been used to detect several features with important practical applications, like heart behavior in patients~\cite{Chang14, Valenza14}, turbulence in solar winds~\cite{Weck15} and atmospheric flows~\cite{Li14} and the evolution of the brain~\cite{Young05}. 

The wide range in which the concept of complexity is applied makes the task of consistently defining a quantitative measure which agrees with intuition in every relevant situation~\cite{Landauer88} a very difficult one, if possible at all. The most sensible way to proceed is to find the basic requirements which are common to a large body of disciplines and try to generalise the obtained model considering the requirements of other fields. Unanimity in this case is probably impossible to achieve, but a compromise can usually be reached.
 
Attempts to define quantitative measures of complexity have appeared especially in physical and mathematical applications. The great majority of proposed measures can be grouped in two main classes which differ by the way they classify random structures. Here, these two classes are going to be called \emph{Type-R} and \emph{Type-S} measures.

Type-R measures define complexity as a measure of randomness in the structure of the considered object or, equivalently, to a lower degree of compressibility of it. The structure of an object is defined to be its description by a certain code (language). A 3-dimensional image of an object, for instance, can be considered as a description of that object by using symbols corresponding to colored volume cells with a positional ``grammar''. Compressibility itself, on the other hand, is a measure of \emph{exact} symmetry breaking as more symmetric objects require less information to be described or reproduced. Algorithmic complexity~\cite{Li97, Zurek89} and Shannon's entropy~\cite{Cover06} both belong to this class. Several Type-R measures are simply variations of the usual entropy definition adapted to different kinds of structures~\cite{Bandt02,Valenza14}. This direct relationship is due to the fact that the more uniform a distribution associated to the description of an object is, the higher is its entropy. 
Therefore, more random objects are naturally more entropic. 

An issue known for a long time \cite{Hogg85} is that the characterisation of complexity purely by randomness is not completely satisfactory in many cases. It has been argued that this choice is physically counter intuitive when applied to several natural complex systems. It is sensible to attribute to biological organisms, for instance, a higher degree of complexity while they are in a structured ``live configuration'' rather than when they attain a deteriorated random ``dead'' one.

Several alternative complexities addressing this matter have been proposed \cite{Bennett86, Grassberger86, Lloyd88, Crutchfield89, Lopez95, Feldman08, Galla12, Aaronson14}. These measures belong to the Type-S class. Like the Type-R ones, they also consider statistical features of either the object or representative ensembles of similar objects, but differently from the latter they rely on averaging out smaller scales, a practice known as 'coarse graining', blurring finer details of the structure which are considered to be not important for the application at hand. The reasoning behind this procedure is that complexity is a concept that depends on the resolution scale of the observer. Usually, the coarse graining scale is chosen somewhat \emph{ad hoc} by appealing to physical considerations about the problem. 

An exception worth mentioning among the many existing measures of complexity is that introduced by Barbour et al. \cite{Barbour14} in the study of gravitating systems of point masses in shape dynamics. Their complexity is defined by purely dynamical quantities and does not seem to belong to any of the two major classes described above. It does not take into consideration statistical features of an ensemble of objects and also does not use any coarse graining procedure. It seems to bear some relation with entropy, but the nature of it is still unclear.  

An ideal measure of complexity would probably need to accommodate features of the two major classes, being based on some sort of general unifying principle allowing a clear path to further generalisation. The present work introduces a complexity measure that addresses these issues by relying on such a new principle - it measures the amount of symmetry which is broken by the description of the object \emph{on average}. This measure can be directly computed from the object being studied, does not use an arbitrary coarse graining procedure and can be generalised straightforwardly for the case of continuous structures. 

In this work we focus on spatial symmetries, using the principle of broken average symmetry to measure the complexity of spatial distributions of points in a square grid. A generalisation to other symmetries needs to take into consideration what is the relevant information that implies complexity in a given scenario. The fundamental guiding principle we are using can then be used for extending this measure to these other symmetries once they are identified. 

In the next section (Sec.~\ref{section:Comps}) a more detailed overview of the currently used complexity measures and how each one deals with randomness is given. Sec.~\ref{section:AS} then discuss the idea of average symmetry and introduce our complexity measure. The following section (sec.~\ref{section:SSS}) shows that the proposed measure can be successfully applied to the case of small scale structures and illustrates it by calculating the complexity of tetrominoes. Sec.~\ref{section:LSS} applies the measure to large scale structures, which includes simple repetitive and self-similar complex patterns. An application of the measure to a satellite image of a fractal river basing is provided as a practical illustration. Finally, conclusions and discussions about the results of this work and future developments are presented on sec.~\ref{section:Conc}. 

% ===================
% Complexity Measures
% ===================
\section{Complexity Measures}
\label{section:Comps}

Let us begin by characterising the Type-R complexities. The most well known of them is Kolmogorov or Algorithmic Complexity (AC). AC defines complexity as the length of the smallest program capable of reproducing an object. Although this definition is clearly language-dependent, it can be shown that different languages will give rise to values of AC that are different by a constant that depends only on the language~\cite{Li97}. In fact, this constant is nothing more than the length of a program that translates from that language to a reference one.

The idea embodied in AC is that more complex objects are more difficult to describe. As any object can be described by simply presenting the object itself, the most complex objects would have descriptions which are nothing but the objects themselves. This measure is connected with compressibility in a straightforward way. Describing an object using less bits than contained in the object itself amounts to compressing it. AC associates complexity to randomness by definition as within its framework a random object is described as that which is incompressible, having no patterns that can be used for creating a optimised description of it and, therefore, being as complex as possible for its size. The justification for using AC as a measure also of randomness is that, having no patterns in its description precludes predictability, which fits well with the intuitive idea of randomness.

As deep and important as AC might be, it turns out that it is uncomputable, although it is bounded by Shannon's entropy which is also a Type-R complexity measure. In fact, there are many bounds that can be found relating the several Type-R complexities~\cite{Aaronson14}. They all share the monotonic relation with respect to randomness presented by AC, which ends up being their most debatable feature. It is usually argued that, while it might be more complicated to describe a random set of points in space than an organised one, too much disorder renders all detailed descriptions of a system useless from a physical point of view to the point which the word `complex' loses its intuitive meaning~\cite{Hogg85}. 

A prototypical example of a physical situation in which this becomes evident is in second order phase transitions~\cite{Yeomans92}, like the Curie point in magnetic systems and type-II superconductivity. The behaviour is universal and can be appreciated in the simplest case of a 2-dimensional Ising model. At zero temperature, the probability distribution concentrates on the ground state, a clearly very simple situation. On the other extreme, that of infinite temperature, all spin configurations contribute equally to the equilibrium state. This situation is not particularly more interesting than the one at zero temperature in general in the sense that the physical behaviour is quite trivial. However, at the critical temperature, where the phase transition from a paramagnetic to a magnetic system occurs, the system is in an extremely interesting self-similar fractal configuration in which all fluctuations scales contribute. 

Right after the problem became widely known, Grassberger~\cite{Grassberger86} suggested a solution that was capable of capturing the higher complexity of the critical point. This complexity became known as \emph{correlation complexity} (CC), and was the first Type-S measure to be proposed. It was based on calculating the entropy of the distribution of different patterns of blocks with a given size and averaging over all sizes with a certain weight. Simulations can show that CC attributes zero complexity to bot zero and infinite temperature configurations of the Ising model, while identifying the critical transition point as the most complex configuration compare with that of other temperatures.   

Others Type-S measures soon appeared. A noteworthy approach was proposed by Crutchfield and Young~\cite{Crutchfield89} and named \emph{statistical complexity}. It relies on a general procedure based on the framework of computational mechanics to describe an object using a dynamical process represented by a finite state machine called an $\epsilon$-machine. An entropy can then be easily defined for every $\epsilon$-machine and this gives the statistical complexity. When applied to physical systems, like the Ising model, it also attributes zero complexity to the extreme temperatures and identify the highest complexity of the critical state~\cite{Crutchfield12}.

Statistical complexity is one possible general approach to complexity, but its intuitive meaning, although well-founded in solid information theoretical principles, is not that clear from a physical point of view. It would be useful to have a possibly more general principle to guide our characterisation of complexity. That does not mean that it would render the former inappropriate, but would maybe point out to more general and encompassing foundations.  

The complexity measure introduced in this work is also a Type-S complexity. Like the other members of this class, it captures the physically expected behaviour of a measure of complexity, but it is obtained from a new powerful and unifying fundamental principle not present in others: that complexity is a measure of the \emph{average} amount of symmetry broken by an object. Very simple objects have great exact symmetry, while an object that can be represented by a highly entropic distribution will have a high degree of symmetry \emph{on average}.

When applied to spatial structures, this principle suggests that one should consider the symmetries of the similarity group of an object - translations, rotations and rescalings. These three completely characterise a physical shape of an object. In the following section we will explicitly find the suggested complexity for this case, obtaining a quantity that is easily computable by a fully parallelisable algorithm and does not rely on coarse graining.

Because it is derived from a solid guiding principle, paths to generalise the new measure, including continuous cases, can be readily identified. Although this measure, like statistical and correlation complexities, attributes zero complexity for homogeneous spaces, it however gives to structured asymmetric objects higher complexity than random ones by detecting that the latter are symmetric \emph{on average}, even if completely breaking exact spatial symmetry. On average, they look the same from every position, in all directions and at all scales. While coarse graining might also detect some average symmetry, it does so by neglecting the object's fine structure. Our measure, on the other hand, does not blur small scale features.

In the following section we will introduce an explicit expression for our complexity for the case of the similarity group and show that the complexity attributed to the spatial structure of several different configurations gives sensible results and can be useful to identify complex patterns in a wide range of applications.

% ================
% Average Symmetry
% ================
\section{Average Symmetry}
\label{section:AS}

A transformation is considered a symmetry of an object if the transformed object retains some characteristic of the original one. Symmetries allow for an object to be reconstructed using the information contained in a smaller part of it together with the knowledge about the specific kind of symmetry it obeys. Rotation symmetry, for instance, allows one to massively compress the description of a circle to one single number: its radius size.  

The concept of average symmetry is reached by relaxing the requirement of exact invariance of an object's property. It is substituted by the requirement that, if the transformation is applied to each element of an ensemble of equally prepared objects, averaging the final objects within this ensemble results in an average object which retains that property even if neither of the individuals in the ensemble does.

For the sake of simplicity, in the present work we will consider discrete structures formed by occupied and non-occupied cells in a square (mostly often 2-dimensional) grid. Later on we will indicate how to generalise it for the continuous case, but we will not enter into the details of this work here. The assumption of discrete grids is not as restrictive as it seems as in most practical applications one will actually work with digital representations of an object which are inherently discrete.   

Consider an $n$-dimensional cubic lattice of side $L$ containing $N=L^n$ sites, or equivalently a cubic grid with $N$ cells, each one either empty or occupied. The finite size of the box will obviously result in finite size effects that will cause fluctuations in the statistical measurements, but one expects these effects to become unimportant as $L$ becomes larger.

Examples of systems that fit this description are the lattice gas and spin systems~\cite{Baxter07}. For the latter, an occupied site can be interpreted as a spin up particle, while a non-occupied one would correspond to a spin down. 

Centered at each site $i$ of the lattice, we consider a cube of edge size $2r+1$, which we call a `scanning ball' of radius $r$. We define the \emph{mass} $\mu_r(i)$ on the surface of this ball as the number of occupied sites on it. We then obtain the normalised mass distribution $\lambda$ at radius $r$ by creating a histogram of the mass values $m$ considering all balls
\begin{equation}
  \lambda(m|r) = \frac1N \sum_{i=1}^N \delta\prs{m,\mu_r(i)},
\end{equation} 
where $\delta$ is a Kroenecker delta. For practical purposes, we consider periodic boundary conditions for the lattices. The complexity $A(L)$ of a certain configuration of sites is defined as the entropy of this distribution averaged from radius 0 to a maximum value $R$ 
\begin{equation}
  A(L) = -\frac1{R+1}\sum_{r=0}^R \sum_m \lambda(m|r) \ln \lambda(m|r),   
\end{equation} 
where $L$ is the system size, and $R$ is the maximum ball radius. By the definition above it is clear that the configurations with all empty sites exchanged by occupied ones and vice-versa have the same complexity. In fact, the difference between the two type is just a question of convention as the entropy is completely symmetric by their exchange. More than that, one can generalise it to configurations in which each site has more than two states which can be labelled arbitrarily. The change of labels, as it should be, is physically meaningless. 

In the case of binary states for each site, when the box is completely full, the mass distribution for any radius concentrates at one single value - the surface area of the ball - and its complexity is trivially zero. The other extreme is a random configuration with each site occupied with probability $p$. For simplicity, let us analyse this latter case in two dimensions, the generalisation to more dimensions is straightforward. The total number of sites (occupied + non-occupied) at radius $r$ is then $8r$. If each site is occupied with probability $p$, the  mass distribution becomes the binomial
\begin{equation}
  \lambda(m|r)=\binom{8r}{m}p^m(1-p)^{8r-m},
  \label{eq:lbd}
\end{equation}
the leading term of its entropy being proportional to the logarithm of its variance $8rp(1-p)$ (see appendix \ref{subsec:BinD}). For large $L$ and the natural choice $R\propto L$, one finds the scale-free result 
\begin{equation}
  \sC(L) = A(L)/\ln L \to 1/2,
\end{equation}
where we are introducing the symbol $\sC$ to indicate the scaled version of the complexity, which turns out to be the most useful in the majority of applications. 

An analogy with fluids can be illustrative. Different occupancy probabilities $p$ can be thought of mass distributions which are homogeneous on average, but with different densities. In the same way as the density of a fluid would be intuitively irrelevant for defining a measure of complexity for it, the exact value of $p$ does not affect the value of our measure. It is then clear that we are able, by using the proposed measure, to discriminate deterministic from \emph{average} homogeneity without focusing on the details of the latter, a property which is not present in other Type-S complexities.   

The shape of the scanning ball is a delicate subject. Tests indicate that, for large boxes, relative values of complexity between different configurations are consistent. However, one has to be careful when considering the appropriate symmetries deemed important for the problem at hand. By choosing a cube, for instance, we are bound to explore only a cubic rotational symmetry of the considered configurations. This is an issue that should be dealt with according to the features of each specific problem. For our purposes in this work, the cubic symmetry of the ball will be enough.  

% ===============================
% Section: Small Scale Structures
% ===============================
\section{Small Scale Structures}
\label{section:SSS}

Let us apply the measure of complexity defined in the previous section to configurations comprised of a small number of points. Clearly, strong fluctuations and anomalies related to the parity of the number of points and their distances are likely to appear in discrete spaces, but these problems should lose relevance either in the limit of a large number of such points or of large systems. The case of one isolated point in a 2-dimensional grid can be solved analytically. The generalisation to more dimensions is straightforward. We restrict the radius of the scanning ball to run from zero to the largest integer less than $L/2$. 
 
The complexity of this system can be easily obtained analytically. All sites of the lattice will have boundaries with zero mass except those located exactly at a square of size $R$ around the occupied site. This amounts for $8r$ points with mass $m=1$ and $L^2-8r$ with $m=0$. The detailed calculation of the scaled complexity is given in appendix \ref{subsec:SingP} and gives, for large $L$,  
\begin{equation}
  \sC(L) \sim \frac1{L}.
  \label{equation:SPC}
\end{equation}

The inverse relation between the scaled complexity and the linear size of the system can be understood by indicating that the contribution of one single isolated point of a extremely large system for its complexity is negligible, a result that agrees with ones intuition about complexity. It is clear that for a large system, adding more points will increase its complexity until their number starts to become comparable with the volume of the system itself. But it is not only the number of points which will affect the complexity of the structure, but also their spatial structure. We can analyse this effect by considering different arrangements of the same number of points in a box. 

\begin{figure}
  \centering
  \includegraphics[width=12cm]{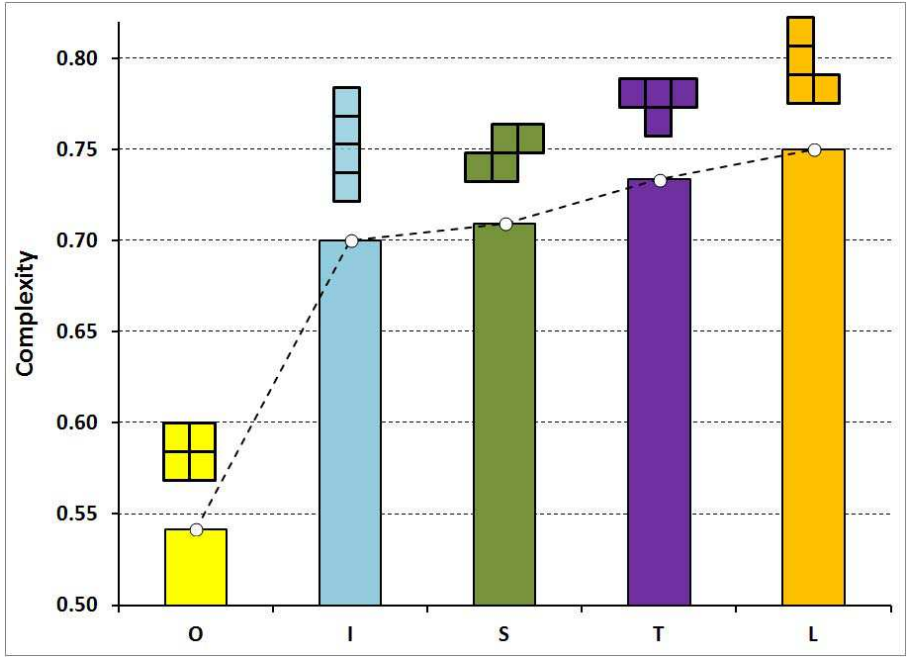}
  \caption{The figure shows the complexity of the five tetrominoes which differ in their symmetry (box size $L=21$). The missing two are the mirror images of L and S, called respectively J and Z, and give the same complexity. The O-piece is the most symmetric and, therefore, is measured as the least complex. The L-piece clearly breaks more symmetries than any other and, consistently, has larger complexity. It is also interesting to notice how close the complexities of I and S are, which is not surprising giving their very similar shapes.}
  \label{fig:4pcomp}
\end{figure}

A very convenient grouping of simple structures is given by \emph{polyominoes} \cite{Golomb94}. These are classes of a fixed number of $p$ adjacent cells in a 2-dimensional grid. The name is a generalization of the word \emph{domino}, which is the special case of $p=2$. As the scale remains the same within a class, symmetry breaking can be clearly visualised. We analyse the \emph{tetrominoes}, polyominoes with four cells popularised by the computer game \emph{Tetris}. There are seven tetrominoes labelled by the letters O, I, S, Z, T, L and J, but only five differ in their symmetries (Z and J are, respectively, mirror images of S and L). By calculating their (not-scaled) complexities in a box with $L=21$ (fig.\ref{fig:4pcomp}) we clearly see that the more symmetric the piece, the less complex it is. The possibility of studying small structures is an important feature of our measure. Reliance on coarse graining would require an artificial fine-tuning of the graining scale. 

% Complexity of Tetris pieces

% ===============================
% Section: Large Scale Structures
% ===============================
\section{Large Scale Structures}
\label{section:LSS}

\begin{figure}
  \centering
  \includegraphics[width=12cm]{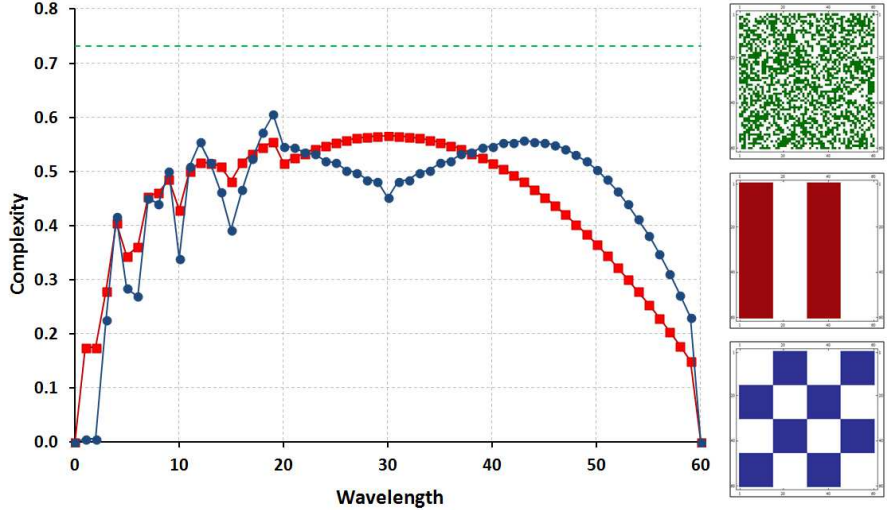}
  \caption{Compared scaled complexities $\sC(L)$ of stripped (squares, online red) and checkerboard (circles, online blue) configurations for different wavelengths. The value of the complexity for a uniformly random configuration (dashed, online green) is also shown for the sake of reference. The pictures on the right show a random configuration (top, online green) and a stripped (middle, online red) and checkerboard (bottom, online blue) configurations of wavelength 15 for illustration.} 
  \label{fig:patcom}
\end{figure}

The proposed complexity measure, although designed to capture average symmetry breaking, must also be able to give sensible results when only exact symmetry breaking is present. This behavior can be analysed by focusing on 2-dimensional patterns whose masses scale with the size of the box $L$. A convenient comparison can be made by using the two patterns that we will call the \emph{stripped} and the \emph{checkerboard} configurations shown on the right side of fig.\ref{fig:patcom} (middle and bottom respectively, red and blue online). Their scaled complexity $\sC(L)$ for different strip/square linear sizes, which we call their \emph{wavelengths}, are shown on the left side of fig.\ref{fig:patcom}. 

All repetitive configurations are less complex than the random case, which has been included for reference in the graph on the right of fig.\ref{fig:patcom} (dashed, online green). This lower complexity comes from the fact that, being exactly symmetric, the patterns contribute for a lower entropy of the mass distribution. Also, there is only one significant scale for these patterns which depends on the wavelength. The checkerboard, for instance, has near-zero complexity for a wavelength of one cell as any ball with the same radius (except zero) has the same mass on its surface. The complexity is locally lower when wavelengths are divisors of $L$, as one can fit an integer number of strips/squares, guaranteeing a higher \emph{exact} symmetry.  

% Complexity of large spatial patterns

Given the above results for the stripped and checkerboard patterns, we would like to compare their complexity with that of structures with broken symmetries, but which are self-similar, like fractals. One would expect intuitively that the latter should be more complex than the former. In particular, self-similar structures should also be more complex than random structures. 

\begin{figure}
  \centering
  \includegraphics[width=14cm]{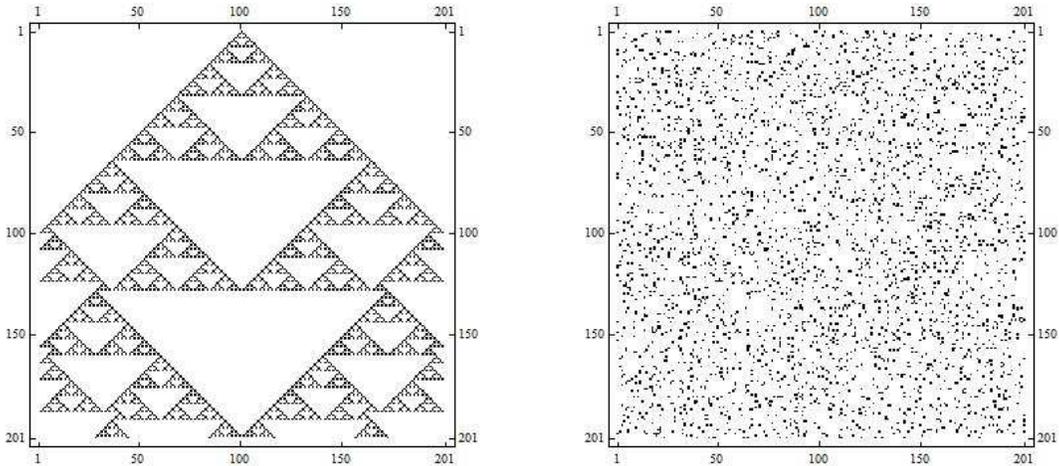}
  \caption{Truncated version of Sierpinski gasket generated by cellular automaton rule number 90 (left) and the same
           sttructure after the occupied sites have been randomly shuffled (right).}
  \label{fig:Sg}
\end{figure}

A convenient way to generate self-similar structures is to use cellular automata. We use Wolfram's rule number 90 \cite{Wolfram02} to generate a truncated Sierpinski gasket. The obtained structure for $L=201$ is given in fig. \ref{fig:Sg} together with the same structure after the occupied cells are randomly shuffled.

\begin{figure}
  \centering
  \includegraphics[width=12cm]{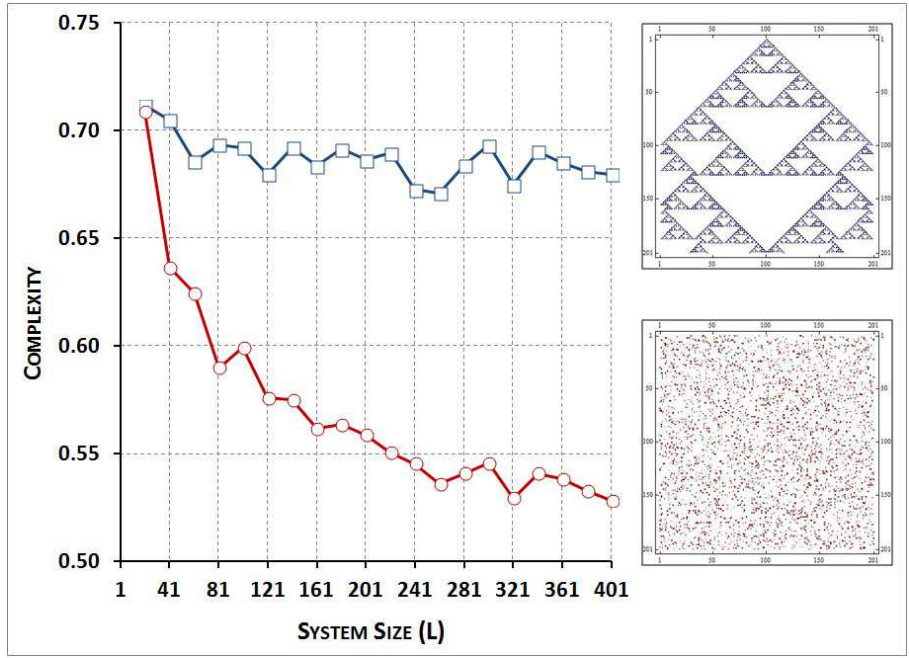}
  \caption{Comparison between the complexity of the Sierpinksi gasket (upper points) and of a random shuffling of its cells (lower points). One can clearly see that the complexity of the fractal structure is always larger.}
  \label{fig:siercomp}
\end{figure}

Fig. \ref{fig:siercomp} shows a comparison of the complexity of the Sierpinski gasket against the shuffled structure for different box sizes showing the expected difference in complexity. Notice that this difference increases with the system size. That is a result of the fact that, at small box sizes, the size of the structure generated by rule 90 becomes too small to show any relevant fractal patterns.

\begin{figure}
  \centering
  \includegraphics[width=12cm]{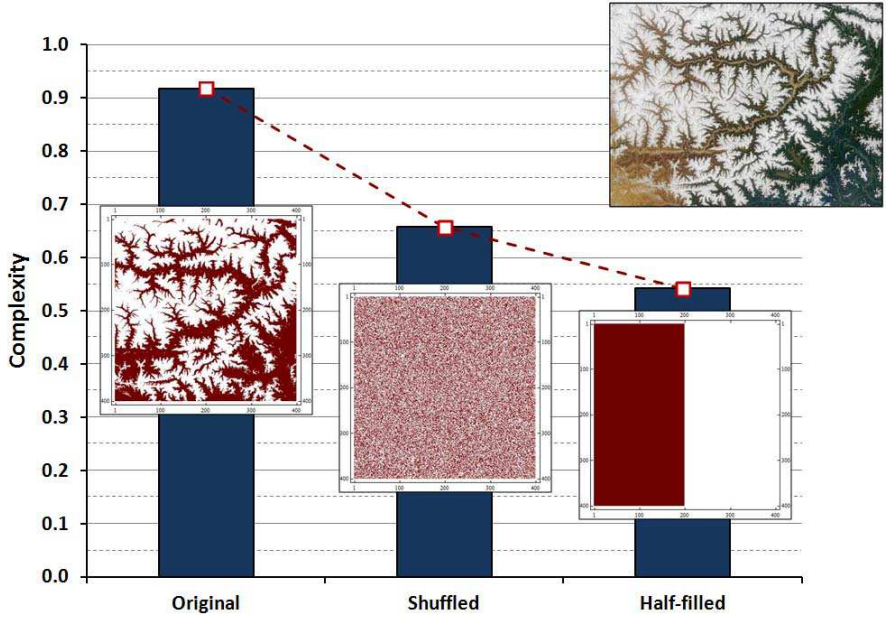}
  \caption{Comparison between the scaled complexity of a natural fractal pattern (left bar), a shuffled configuration of it (middle bar) and a half-filled configuration of same dimensions (right bar). The difference in complexity of can be clearly seen in the picture. The upper right inset shows the original picture, an image of the Yarlung Tsangpo River, China, taken by NASA's Terra satellite.}
  \label{fig:fracshuf}
\end{figure}

To illustrate how the proposed measure behaves in practical situations, we used an image of the Yarlung Tsangpo River, China, taken by NASA's Terra satellite. River basins are one of the most common natural occurrences of approximately fractal patterns. The original picture (fig. \ref{fig:fracshuf}, top right) was 1079$\times$802 pixels with a resolution of 72dpi. It was rescaled to 538$\times$400 pixels in an ordinary graphic editor and had its contrast adjusted to a maximum in order to become purely monochrome. Finally, it was cropped to a 400$\times$400 pixels square and turned into a binary square matrix with $L=400$. Its scaled complexity (fig. \ref{fig:fracshuf}, left) was then calculated and compared to that of the same image with its occupied sites shuffled randomly (fig. \ref{fig:fracshuf}, middle), showing the significantly higher complexity of the fractal pattern. A picture of a half-filled box of same size is also presented (fig. \ref{fig:fracshuf}, right) to show its even lower complexity. 

% Fractal geological feature

% ======================
% Section: Gas Automaton
% ======================
\section{Gas Automaton}

We now address how the complexity changes with time during the time evolution of a discrete statistical physics model. We consider a 2-dimensional lattice gas cellular automaton similar to that used in a paper by Aaronson et al.~\cite{Aaronson14}, where it was called the \emph{Coffee Automaton}. We use a box with linear size $L=100$ and an initial configuration in which the gas occupies the whole left half of this box. The automaton simulates the gas expansion until it occupies uniformly the whole available volume.

Because the main objective is to obtain the behavior of its complexity, we chose a very simple automaton rule. In each time step, a site and one of its 8 neighbours are chosen at random uniformly from the whole box. If one of the sites is occupied and the other empty, their status is exchanged, otherwise they remain the same.

One would expect that the complexity starts with a low value due to the simple initial condition and ends close to 1/2 (due to the finite size of the box, finite size effects might be significant) for a random final configuration. At intermediate time steps, the complexity should increase to a maximum and then decrease. This picture can be seen as an extreme simplification of the cosmological evolution of matter in our universe~\cite{Aaronson14}. Our present view is that the universe started in a very simple configuration and will end in another simple random one as it runs towards its state of thermal death. However, the intermediate configurations of matter are complex enough to support life-forms with universal computation capabilities like humans. 

\begin{figure}
 \centering
  \includegraphics[width=12cm]{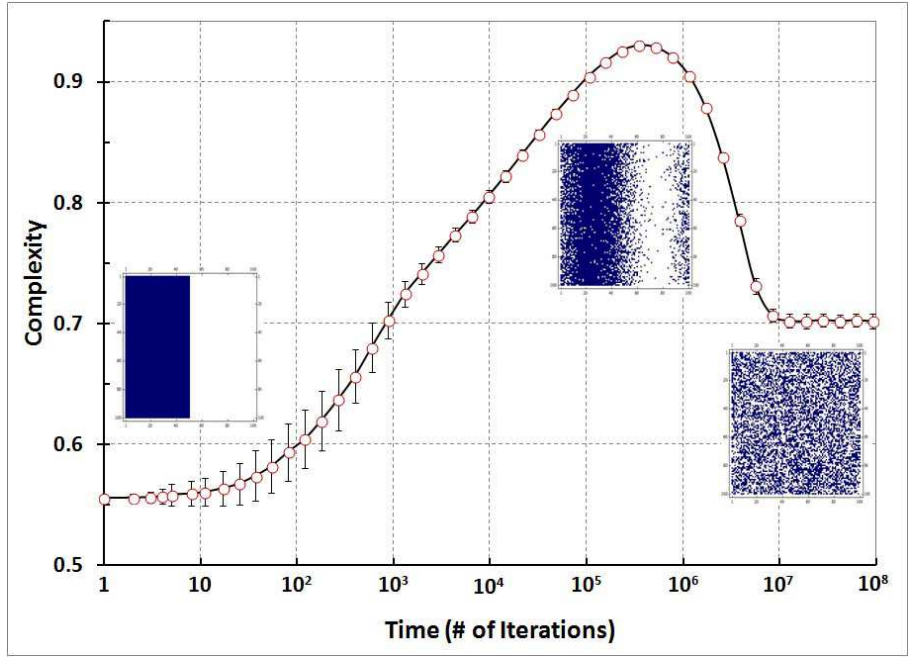}
  \caption{The plot shows the evolution of complexity during the expansion of a 2-dimensional lattice gas which starts confined in the left half of a square box of size $L=100$. Time (as the number of iterations of the automaton rule) is presented in a logarithmic scale. Three snapshots of the gas configuration are also shown respectively at the beginning, the point of maximum complexity and at the end, when the gas settles down to a random equilibrium configuration. Error bars correspond to variances over 100 repetitions of the process.}
  \label{fig:gas}
\end{figure}

Fig. \ref{fig:gas} shows the plot of the scaled complexity averaged over 100 repetitions of the expansion process together with the corresponding error bars. The time scale of the graph is logarithmic due to the fact that, as the gas expands, it takes an increasing number of iterations of the rule in order to change the configuration in a significant way. One can clearly see the initial increase and late decrease in the complexity, with the differences for the initial and final configurations reflecting their exact and average symmetry breaking respectively.  

% Complexity of gas mixing

% ===========
% Conclusions
% ===========
\section{Conclusions}
\label{section:Conc}

This work introduced a measure of complexity based on a new principle: complexity as a measure of average symmetry breaking. We focused here on spatial symmetries of discrete lattices, probing the homogeneity, isotropy and scale invariance (invariance under the similarity group) of configurations of empty and occupied sites. This was done by defining the complexity as the entropy of the ``mass distribution'' on the surface of moving scanning balls averaged over all scales. Using several different examples, we showed that the results are consistent with intuitive expectations and correctly capture the different behaviors of the studied classes of configurations.

The measure introduced here attributes low complexity both to very simple repetitive structures and to uniformly random ones, while giving higher complexity to intermediate structures. In particular, it classify self-similar and fractal structures as more complex than the former two. Although other complexities also present this behavior, ours differs by, at the same time, capturing also the difference between simple \emph{deterministic} and simple \emph{random} configurations, a feature that is not present in other measures. In addition, due to its simple interpretation, it is easy to generalise to several different situations. The reliance on symmetry breaking also leads us to conjecture that it has a more general formulation which is applicable to general symmetry groups.

The shape of the scanning balls used to define the mass distribution has been arbitrarily chosen to be a square, which has the consequence of, instead of full rotation symmetry, only being able to scan for C4 symmetry. A different choice would be to consider the ball as formed by all sites at the same path distance from the central one. In the large system limit, this should be able to capture spherical symmetry. A more correct choice could be to consider averages over random shapes. This is however a more involved study which we leave for future work.

Although we used in a binary discrete variable at each site (empty/occupied), our scaled complexity $\sC(L)$ can be readily generalised to other kinds of configurations. An immediate one would be for physical situations requiring sites with additional properties, like charges, spin or color~\cite{Mulet02}. This can be trivially included by considering a sample space of the appropriate size when calculating the probabilities. For instance, while a binary situation is appropriate to consider spin configurations like in a simple Ising model, a spin-1 model requires the probabilities for the three spin values 0, 1 and -1. 

A different extension would be to consider the classification of complexity of more general network topologies than a regular lattice as used here, a problem with several practical applications (\cite{Wainrib13}, for instance). By measuring the number of sites at fixed path-distances around an initial one, we would obtain the result that \emph{regular} graphs have zero complexity. This would of course only measure the \emph{intrinsic} complexity of the graph, the complexity calculated by an observer living inside of it. A different result would be obtained by considering the graph embedded in a larger network, which we would call the \emph{extrinsic} complexity of the graph. We are currently studying the properties of these two different network complexities and the results will be published elsewhere. 

Finally, the extension for continuous configurations can be done by considering a finite volume $V$ in a $d$-dimensional space. The total mass contained in a \emph{spherical shell} of radius $r$ and thickness $\epsilon$ around a point $\vc{x}$ in this volume becomes
\begin{equation}
  \mu_{r,\epsilon} (\vc{x})= \int_r^{r+\epsilon} dr' r'^{d-1} \int d\Omega \; \rho(\vc{x}+\vc{r}'),
\end{equation} 
where $d\Omega$ is the angular element of the integration and $\rho$ is the mass density at each point of the volume. One then uses this function to calculate a distribution of masses
\begin{equation}
  \lambda(m|r,\epsilon) = \frac1V \int dV\; \delta\prs{m-\mu_{r,\epsilon}(\vc{x})}.
\end{equation}

The complexity is then the average differential entropy. It has a different range from the discrete case as it diverges when the distribution peaks at a single point and might have negative values. When the distribution is uniform over a volume, the entropy becomes simply the logarithm of the volume and diverges as it increases. Although one faces these divergences, the overall behaviour is still consistent. A conformal mapping like that provided by the hyperbolic tangent could be used in this case, but we leave this study for a future work.

Through a sequence of symmetry breakings our universe changed from a simple homogeneous state to the complex structure we observe today. But while this process holds the key to create the present diversity of shape and function, too much of it can destroy any interesting feature separating complexity from chaos. True complexity is found in between total order and total disorder. We have shown here that we can consistently characterise this middle point by measuring the average symmetry of an object. Symmetry is a strong and general concept that pervades every discipline, from arts to science, much like complexity. If one seeks a unified framework for complexity, a measure that can be used across disciplines is of utmost importance. It has to be easy to calculate, consistent, readily generalisable to new phenomena and based on solid principles. We believe that the measure presented here has all these properties.

% =========================
% Section: Acknowledgements
% =========================
\section*{Acknowledgments}

I would like to thank Julian Barbour, Juan Neirotti and David Saad for useful discussions and suggestions.

% ============
% Bibliography
% ============

\bibliographystyle{apsrev4-1}
\bibliography{complex}

\appendix

% =================
% Section: Appendix
% =================
\section{Complexity for Special Cases}

% ---------------------------------
% Subsection: Binomial Distribution
% ---------------------------------
\subsection{Binomial Distribution}
\label{subsec:BinD}

The binomial distribution given by equation (\ref{eq:lbd})
\begin{equation}
  \lambda(m|r)=\binom{8r}{m}p^m(1-p)^{8r-m},
\end{equation}
in the limit of large $r$ approaches the Gaussian
\begin{equation}
  \lambda(m|r) \approx \frac1{\sqrt{2\pi\sigma^2}}\exp{\frac{(x-\mu)^2}{2\sigma^2}},
\end{equation}
with $\mu=8rp$ and $\sigma^2=8rp(1-p)$.

For a large 2-dimensional box, the leading contribution for the scaled complexity can be calculated using the entropy of this Gaussian distribution
\begin{equation}
  \sC(L) = \frac1{R} \sum_{r=1}^R \frac12\ln 2\pi\sigma^2,
\end{equation}
where the $r=0$ contribution was neglected as it disappears for large $R$. In this limit, the only surviving term is the logarithm of $r$, which gives
\begin{equation}
  \begin{split}
    \sC(L) &=       \frac1{2R\ln L} \sum_{r=1}^R \ln r\\
           &=       \frac1{2R\ln L} \ln R!\\
           &\approx \frac1{2R\ln L} (R\ln R-R),
  \end{split}
\end{equation}
which gives the result 1/2 for $L\rightarrow\infty$ with $R\propto L$ independently of $p$.

% ------------------------
% Subsection: Single Point
% ------------------------
\subsection{Single Point}
\label{subsec:SingP}

For one single point in a lattice of size $L$, the mass distribution becomes
\begin{equation}
  \lambda(m|r) = \frac1{L^2} \col{(L^2-8r)\delta\prs{m,0}+8r\delta\prs{m,1}},
\end{equation} 
which gives
\begin{equation}
    A(L) = \frac1{R+1}\sum_{r=0}^R \chs{2\ln L - \frac1{L^2}\col{(L^2-8r)\ln (L^2-8r)+8r\ln 8r}}. 
\end{equation}

The first term is simply $2\ln L$. The next two terms can be approximated by an integral when $L$ is large with the results
\begin{align}
  \frac1{L^2(R+1)}\sum_{r=0}^R (L^2-8r)\ln (L^2-8r) &\approx \frac1{RL^2}\int_0^R dx\; (L^2-8x)\ln (L^2-8x)\\ 
                                                    &\sim    2\ln L-4\frac{\ln L}{L}-\frac12+\frac1L,\nonumber\\ 
  \frac1{L^2(R+1)}\sum_{r=0}^R 8r\ln 8r             &\approx \frac1{RL^2}\int_0^R dx\; 8x\ln 8x\\ 
                                                    &\sim    6\frac{\ln 2}{L} +\frac12\frac{\ln L}{L}-\frac1L.\nonumber
\end{align}

Adding all together one obtains 
\begin{equation}
  A(L) \approx 4\frac{\ln L}{L}-6 \frac{\ln 2}L, 
\end{equation}
which gives equation (\ref{equation:SPC}) in the large $L$ limit.

\end{document}